\documentclass[amsmath, a4paper, prb, twocolumn, showpacs]{revtex4}
\usepackage{amssymb}
\usepackage{amsmath}
\usepackage{color}
\usepackage{graphics, graphicx}
\usepackage{bbold}
\usepackage{psfrag}
\usepackage{mathcomp}
\usepackage{subfigure}
\usepackage{dsfont}

\begin{document}
\title{Canted Antiferromagnetic Order of Imbalanced Fermi-Fermi mixtures in Optical Lattices by Dynamical Mean-Field Theory}
\author{Michiel Snoek$^{1}$}
\author{Irakli Titvinidze$^{2}$}
\author{Walter Hofstetter$^{2}$}
\affiliation{$^{1}$
Institute for Theoretical Physics, University of Amsterdam, 1090 GL Amsterdam, The Netherlands 
 \\ $^2$Institut f\"ur Theoretische Physik, Johann Wolfgang Goethe-Universit\"at, 60438 Frankfurt/Main, Germany}


\begin{abstract}
We investigate antiferromagnetic order of repulsively interacting fermionic atoms in an optical lattice by means of Dynamical Mean-Field Theory (DMFT). Special attention is paid to the case of an imbalanced mixture. 
We take into account the presence of an underlying harmonic trap, both in a local density approximation and by performing full Real-Space DMFT calculations. 
We consider the case that the particle density in the trap center is at half filling, leading to an  antiferromagnetic region in the center, surrounded by a Fermi liquid region at the edge. In the case of an imbalanced mixture, the antiferromagnetism is directed perpendicular to the ferromagnetic polarization and canted. We pay special attention to the boundary structure between the antiferromagnetic and the Fermi liquid phase. 
For the moderately strong interactions considered here, no Stoner instability toward a ferromagnetic phase is found. 
Phase separation is only observed for strong imbalance and sufficiently large repulsion.
\end{abstract}

\maketitle

\section{Introduction}
Ultracold atoms in optical lattices provide a versatile laboratory for interacting quantum many body systems \cite{Bloch}. One of the major challenges in this field is the experimental investigation of quantum magnetism in atomic mixtures. Impressive experimental progress in this direction has already been made.
The first important step in experiments with fermionic atoms in optical lattices was the experimental observation of the Fermi surface  \cite{Kohl}.
Recent experiments with bosonic atoms directly observed correlated particle tunneling \cite{Foelling07} and superexchange \cite{Trotzky08}, which are the basic mechanisms underlying quantum antiferromagnetism. Moreover, strong experimental evidence for the fermionic Mott insulator state was obtained, both by the local probe of observing reduced double occupancy \cite{Jordens08} and the global probe of observing a plateau in the cloud size when the system is compressed \cite{Schneider08}. 
A recent experiment in a system of spin-1/2 fermions without optical lattice indicates a Stoner instability toward a ferromagnetic state for strong repulsion \cite{Jo09}.
These are important steps on the way toward realization of strongly correlated many-fermion states \cite{Hofstetter02}. 
Currently the experimental temperatures are still higher than the critical (N\'eel) temperature, below which antiferromagnetic order is predicted to develop \cite{Werner05, Koetsier07}. Most accurate theoretical estimates for the entropy per particle below which long-range antiferromagnetic order is expected yield a value of $S/N \approx \ln(2)/2$ \cite{Leo08, Wessel10}, whereas current experiments reach an average entropy which is still a factor $2$ higher \cite{Jordens10}. 


Ultracold atomic system offer the unique possibility to control the relative densities of the two spin components, as alreday  has been demonstrated in experiments without the presence of an optical lattice \cite{Zwierlein06, Partridge06}.  
Experimentally, the density imbalance is precisely tunable by means of radiofrequency sweeps  \cite{Zwierlein06, Partridge06} and stable due to the suppression of spin-flip scattering processes in cold atomic gases. This realizes an imbalanced spin mixture, in which the $SU(2)$-symmetry is broken by an artificial magnetic field.  When the density of atoms corresponds to one particle per lattice site, the ground state of this system is expected to be a canted antiferromagnet, with antiferromagnetic order characterized by a N\'eel vector directed  perpendicular to the applied field. 
However, experimentally ultracold atom systems are always confined by an external harmonic trapping potential, which leads to an inhomogeneous system.
If the total particle number is sufficiently high, in the center of the trap a region with particle density per site close to one will develop, where antiferromagnetic order is stable at sufficiently low temperatures \cite{Snoek08, Gorelik10}. The edges of the system have lower filling; they are Fermi liquid regions without spin order. If the total particle number is even higher, also in the trap center a Fermi liquid with particle density higher than one or a band insulating state can exist. In that case antiferromagnetic order can be stable in a shell around this Fermi liquid \cite{Snoek08}. 
This poses interesting questions regarding the nature and the stability of spin order, which we will address in this paper by means of (Real-Space) Dynamical Mean-Field Theory. 


These issues have recently also been investigated by other methods. For homogeneous systems described by the hole-doped Hubbard model, both commensurate and incommensurate spin-density-waves have been predicted \cite{incommensurate1, incommensurate2, incommensurate3}. By mapping to an effective spin model, the critical temperature for canted antiferromagnetic order was calculated and topological excitations of imbalanced mixtures were studied \cite{Koetsier09}.
A Hartree-Fock static mean-field theory  for balanced mixtures in a trap predicts 
that antiferromagnetism
can coexist with paramagnetic states in various spatial patterns, for example antiferromagnetism in the center of the trap surrounded by a hole-doped atomic liquid or antiferromagnetism in a ring with a Fermi liquid in the center and at the edge \cite{Andersen}. For imbalanced mixtures, this approach predicts canted order perpendicular to the (artificial) magnetic field up to moderate values of the repulsion \cite{Gottwald09}. Very recently the Hartree-Fock approach has also been applied to larger repulsion: in addition to canted antiferromagnetism,  a critical interaction was found, beyond which the Stoner instability drives a ferromagnetic transition at the edge of the system, where the particle density is lower than half-filling \cite{Wunsch09}. 

 A Real-Space Dynamical Mean-Field (R-DMFT) study of antiferromagnetism in a harmonic trap has also been performed, but so far without allowing for the possibility of canted antiferromagnetic order \cite{Snoek08, Gorelik10}. For the case of an imbalanced mixture, this constraint lead to the prediction of phase separation between the majority component in the center and the minority component at the edge for sufficiently strong repulsive interactions and large values of the imbalance \cite{Snoek08}.

Here we perform a full R-DMFT study, which includes the possibility of canted order. Unlike static Hartree-Fock mean-field theory, DMFT is a non-perturbative method which is reliable both for strong and weak interactions in sufficiently high dimensions. Local correlations are included exactly \cite{Georges1, Georges2, Georges3, Georges4}. R-DMFT thereby takes the inhomogeneity induced by the presence of a harmonic trap into account in a fully consistent way.

For imbalanced systems we indeed observe canted antiferromagnetic order. 
We consider weak to moderately strong interactions, for which no Stoner instability toward spontaneous ferromagnetism is found: in the case of a balanced mixture the wings of the systems are always paramagnetic. 
Only upon applying a finite amount of imbalance, the system gets polarized and ferromagnetic order starts to develop. We generally also do not observe the phase-separation scenario for large imbalance. Instead, the canted antiferromagnetic order allows for a continuous transition between balanced antiferromagnetism order and fully imbalanced ferromagnetic order. Only for large values of the interaction and strong imbalance phase separation occurs.

We compare our full R-DMFT results 
with calculations based on a local density approximation in combination with DMFT, in which the harmonic trap
is incorporated by a spatially varying chemical potential. As for the balanced case \cite{Snoek08, Gorelik10} we find that the total density is well approximated by the local density approximation, but a strong proximity effect is observed for the antiferromagnetic order: 
the staggered antiferromagnetism as obtained by the full R-DMFT calculation
extends to regions where the local density approach predicts a paramagnetic solution.

\section{Model}
Repulsively interacting fermions in a sufficiently deep optical lattice are well described by the single-band Hubbard Hamiltonian in the tight-binding approximation
\begin{equation} \label{Hamiltonian}
\mathcal{H} =-J \hspace{-1mm} \sum_{\langle ij\rangle,\sigma} \hat c^\dag_{i\sigma} \hat c_{j\sigma} +U\sum_{i} \hat n_{i\uparrow} \hat n_{i\downarrow} + \sum_{i\sigma} (V_i - \mu_\sigma) \hat n_{i\sigma},
\end{equation}
where
$\hat n_{i\sigma}= \hat c^\dag_{i\sigma}\hat c_{i\sigma}$, and $ \hat c_{i\sigma}$ ($\hat c^{\dagger}_{i\sigma}$) are fermionic annihilation (creation) operators for an atom with spin $\sigma$ at site $i$, 
$J$ is the hopping amplitude between nearest neighbor sites $\langle ij\rangle$,
$U>0$ is the on-site interaction, $\mu_\sigma$ is the (spin-dependent) chemical potential and $V_i=V_0 r_i^2$ is the harmonic confinement potential. We also define $\bar \mu \equiv \frac{1}{2}(\mu_\uparrow+\mu_\downarrow)$ and $\Delta \mu \equiv \mu_\uparrow-\mu_\downarrow$ as the average chemical potential and difference in chemical potential, respectively. Although $\Delta \mu$ acts as a magnetic field,  experimentally the imbalance is not induced by a physical magnetic field, but by directly controlling the difference in particle number. 
The parameters of this model can be tuned in experiments by  changing the intensity of the optical lattice and via Fesh\-bach resonances \cite{Bloch}. In the following, we take the lattice constant to be $a=1$.

\section{Method}
To obtain the ground state properties of this system, we apply R-DMFT \cite{Dob98, Potthoff99, Tran06, Tran07, Song08, Helmes08, Snoek08, Koga09, Gorelik10, Gorelik10b, Kim10}.
Within R-DMFT the self-energy is taken to be local (which is exact in the infinite-dimensional limit \cite{Metzner1, Metzner2}) but allowed to depend on the lattice site, i.e. $\Sigma_{ij\sigma} (i \omega_n) =\Sigma_{\sigma}^{(i)}(i \omega_n) \delta_{ij}$, where $\delta_{ij}$ is a Kronecker delta. The lattice sites are described by local effective actions, each representing an effective Anderson impurity model, which are coupled via the real-space Dyson equation for the Green's function. Details of the method have been published previously \cite{Snoek08}.

In the present paper we use Exact Diagonalization (ED) \cite{Georges3, Caffarel94, Si94} of the Anderson Hamiltonian to solve the local impurity actions.
Within ED the spectral function is represented by a finite number of delta peaks. Whereas this is sufficient for a faithful representation of the zero-temperature spectral function, we found it to lead to unphysical behavior at finite temperature, especially away from half-filling. Therefore we restrict ourselves in this article to the low-temperature limit and only investigate ground state properties.
The multigrid Hirsch-Fye quantum Monte Carlo method has proven to be a very efficient solver at finite temperatures for a balanced mixture
\cite{Bluemer08, Gorelik10, Gorelik10b}, but canted antiferromagnetic order is probably harder to obtain within this method. Also when using the Numerical Renormalization Group (NRG) method\cite{Wilson75, Bulla08, Snoek08} to solve the Anderson Hamiltonian, it is problematic to describe the canted off-diagonal spin order. 
In contrast, the ED-method we use here is very flexible, which also allows to incorporate off-diagonal canted spin order in a straightforward manner.  However, since $S_z$ in this case is not a good quantum number for the individual spin components, the size of the Hilbert spaces to be diagonalized is significantly enlarged, which leads to far more time-consuming numerics compared to the balanced case.

For very large repulsion, ED can run into unphysical instabilities. Therefore we consider only moderately large ratios of $U/J$ here. 

\begin{figure}
\includegraphics[width =1 \columnwidth]{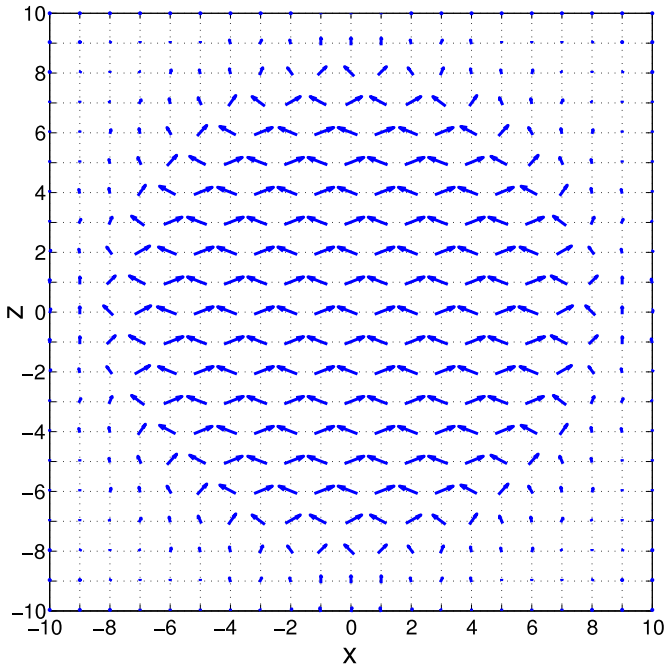}
\caption{(Color online) Real-space spin profile for $U/J = 10$, $\Delta \mu/J = 0.5$ and $V_0/J = 0.05$. 
The vertical components of the arrows symbolize the local magnetization in the $z$-direction $\langle \hat S^z_i \rangle$, while the horizontal components correspond to the (staggered) local magnetization in the $x$-direction $\langle \hat S^x_i \rangle$.
 For all data in this paper we took the filling at the center equal to one, which is induced by the choice of the chemical potential equal to $\bar \mu = U/2$. To obtain the ground state expectation values a small fictitious temperature of $T/J=0.02$ was applied. 
}
\label{RS2d}
\end{figure}

In the practical implementation of ED a small but finite temperature is used to generate the $T=0$ data, 
in order to obtain discrete Matsubara frequencies. 
We chose this temperature equal to $T/J = 0.02$ for the two-dimensional data and $T/J = 0.025$ for three dimensions.

\section{Results}
We now apply the R-DMFT method to spin-$\frac{1}{2}$ fermions in a two-dimensional square lattice and a three-dimensional cubic lattice with harmonic confinement. We focus on the density distributions of the two species $n_{i\sigma} = \langle  \hat n_{i\sigma} \rangle$ ($\sigma = \uparrow, \downarrow$), the total density $n_{i{\rm tot}} = n_{i\uparrow}+n_{i\downarrow}$, and the local spin expectation values $S_i^\alpha = \frac{1}{2} \langle \hat c_{i\beta}^\dagger \sigma_{\beta \gamma}^{\alpha} \hat c_{i\gamma} \rangle$, in which $\sigma^\alpha$ ($\alpha=x,y,x$) are the Pauli matrices. Here we have set $\hbar = 1$.

In the case of a balanced mixture the Hamiltonian is $SU(2)$-symmetric. This means that the staggered magnetization can point in any direction. We have chosen it to point in $x$-direction. In the case of an imbalanced mixture, the $SU(2)$-symmetry is spontaneously broken by the chemical potential difference, which acts as an artificial magnetic field in the $z$-direction. In reaction, the staggered magnetization orders perpendicular to this, i.e. in the $xy$-plane. The remaining $U(1)$ symmetry is also in this case in our calculations
broken by a small initial numerical perturbation, resulting in alignment of the spins along the $x$-direction, such that  $\langle \hat S^y \rangle=0$ in all results presented here.

In all the calculations reported here we have chosen $\bar \mu = U/2$, such that the system is at half-filling at the center.

\begin{figure}
\hspace{-1cm}
\includegraphics[scale = .4]{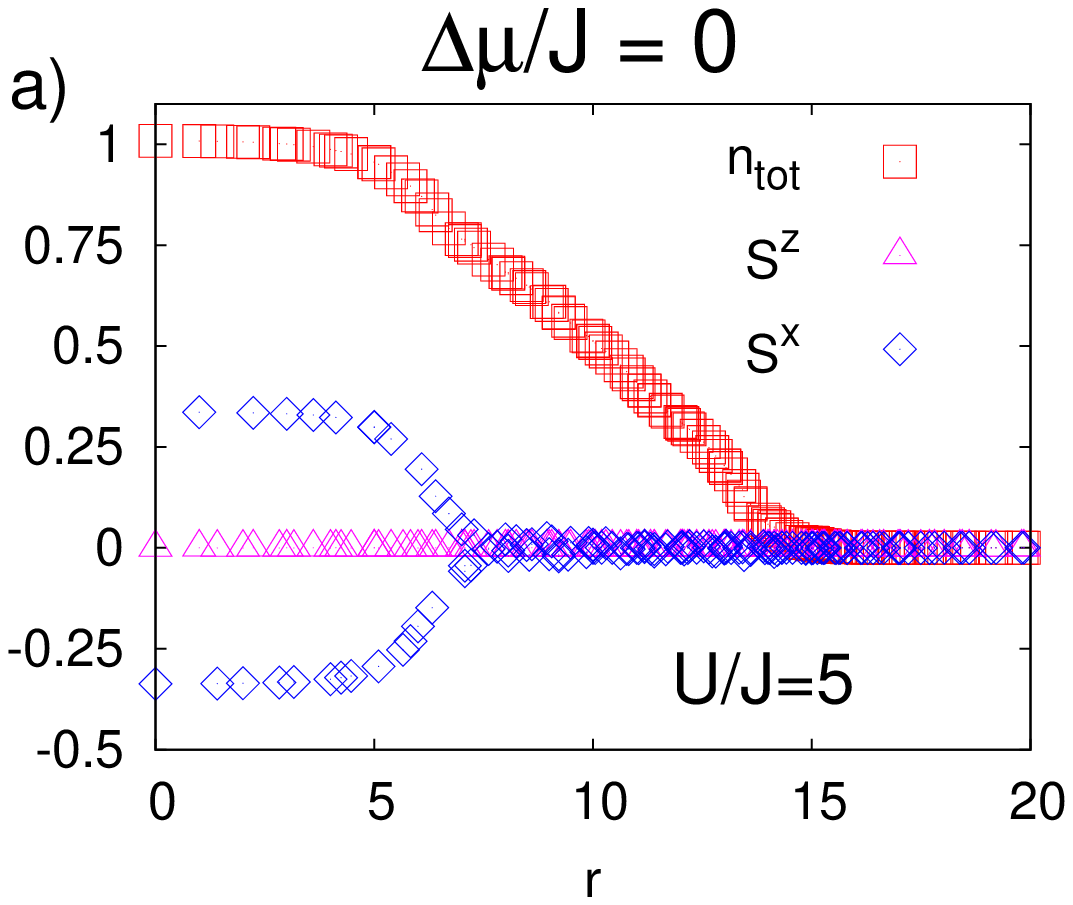}\hspace{-.8cm}
\includegraphics[scale = .4]{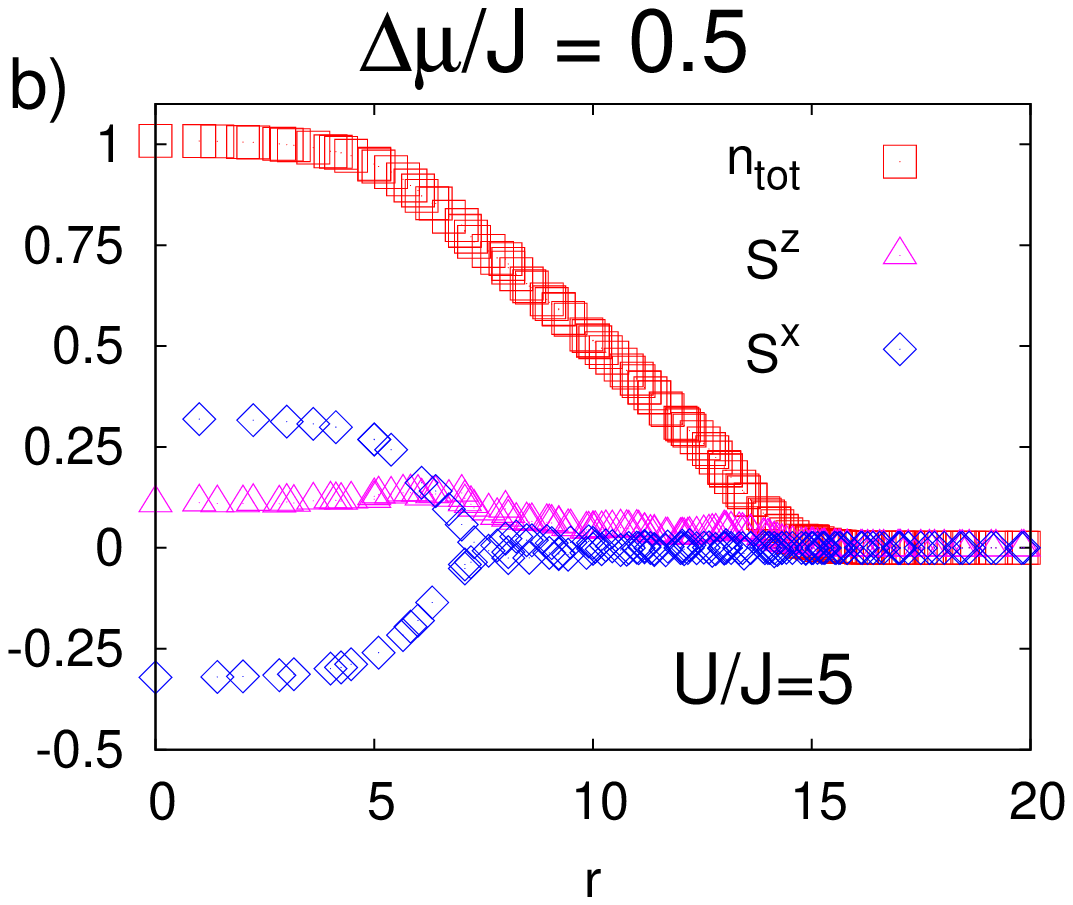}\hspace{-.5cm}

\hspace{-1cm}
\includegraphics[scale = .4]{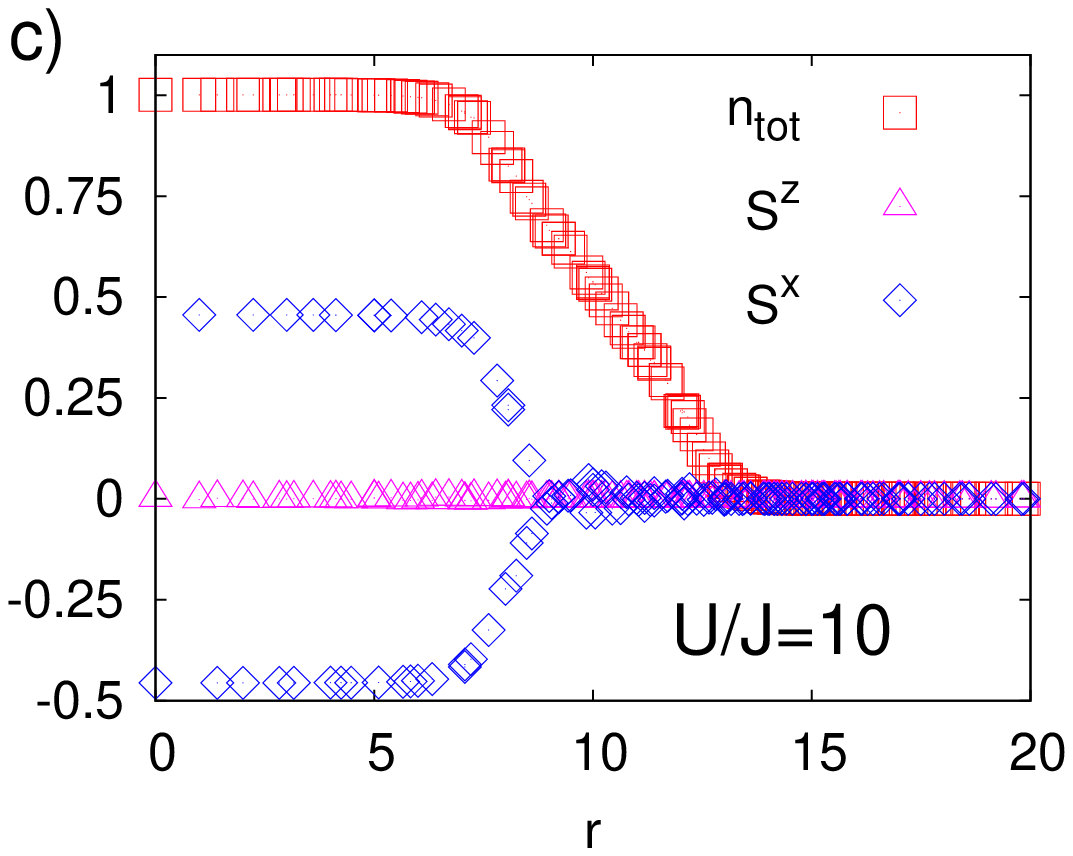}\hspace{-.8cm}
\includegraphics[scale = .4]{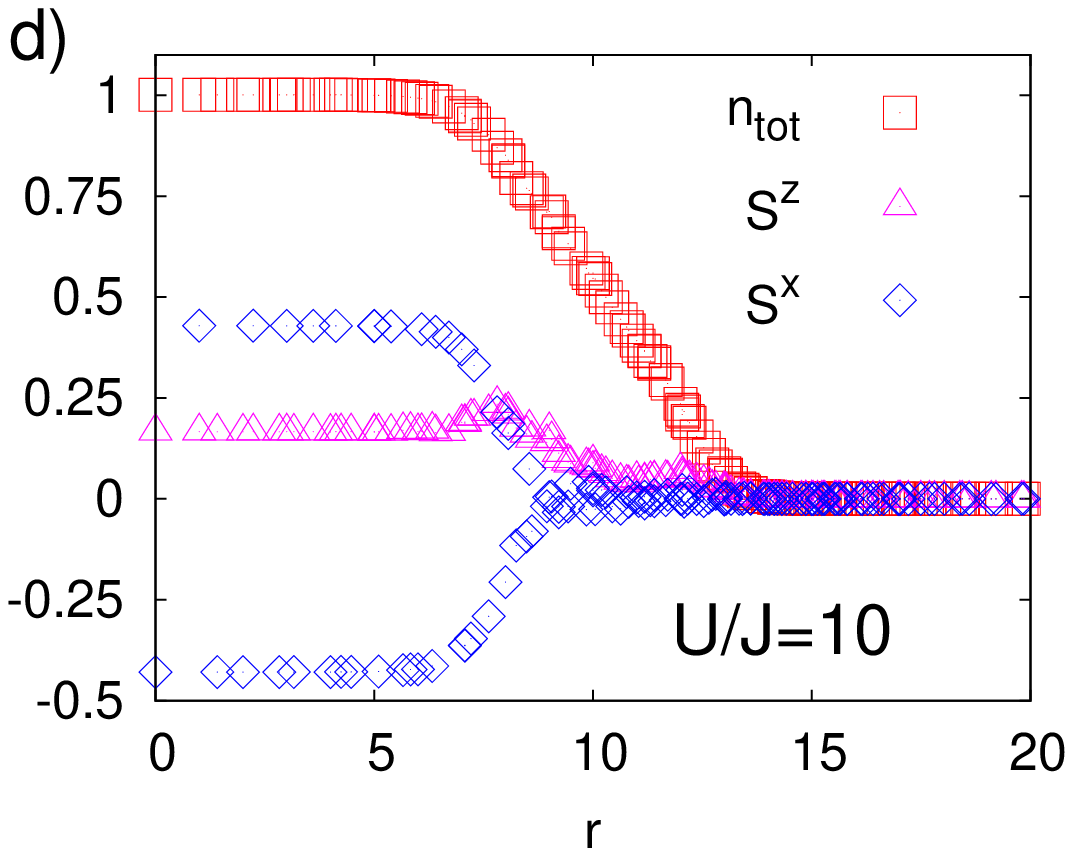}\hspace{-.5cm}

\hspace{-1cm}
\includegraphics[scale = .4]{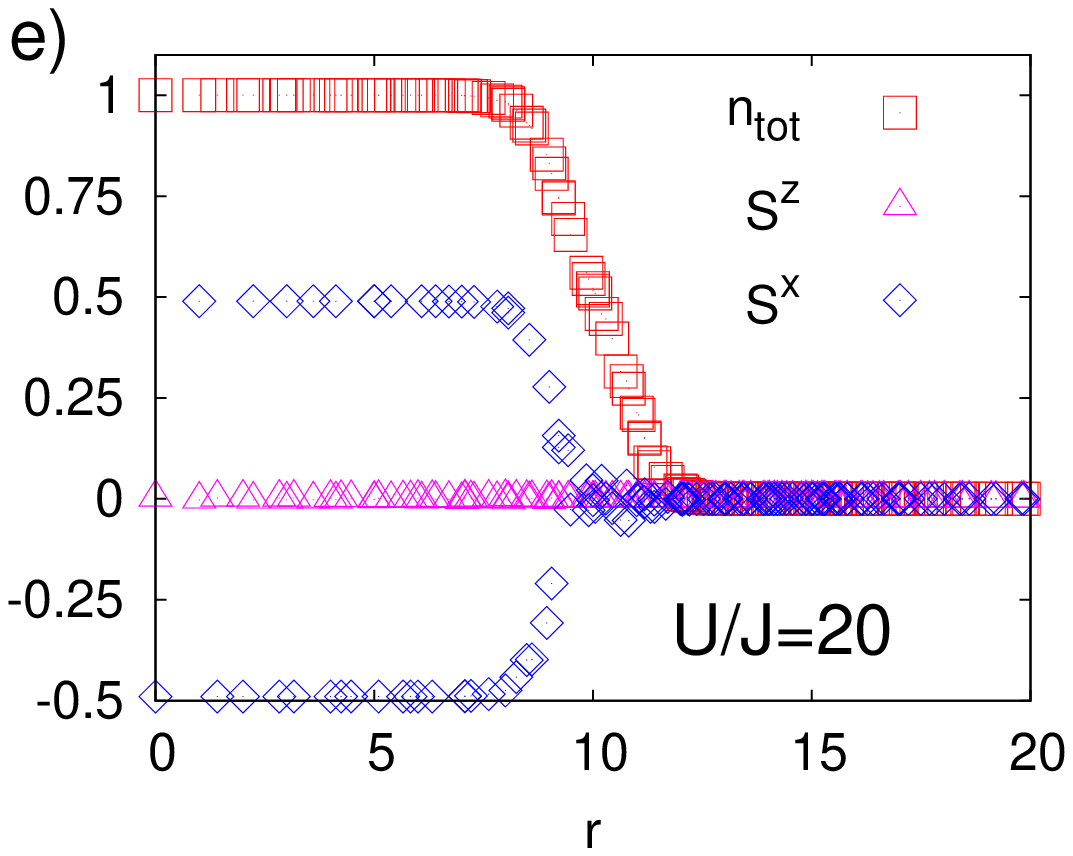}\hspace{-.8cm}
\includegraphics[scale = .4]{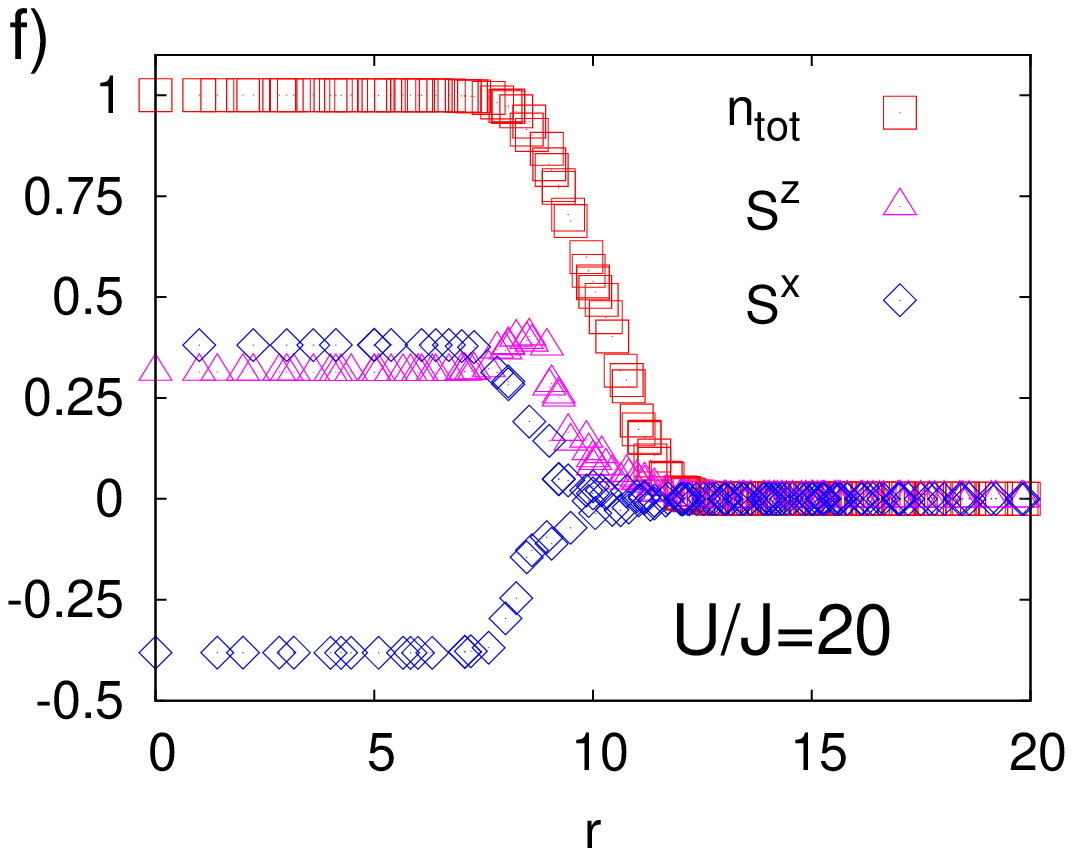}\hspace{-.5cm}

\caption{(Color online) Radial profiles for the ground state of \emph{two-dimensional} Fermi-Fermi mixtures for different values of $U$ and $\Delta \mu$ . Plotted are the total density $\langle \hat n_i \rangle = \langle \hat n_{i\uparrow} + \hat n_{i\downarrow} \rangle$, and the expectation values of the spin in $z$-direction $\langle \hat S^z_i \rangle$ and in $x$-direction $\langle \hat S^x_i \rangle$. The left column is for a balanced mixture ($\Delta \mu/J = 0$), the right column is an imbalanced mixture with $\Delta \mu/J = 0.5$. The other parameters are $\bar \mu = U/2$ and from top to bottom: $U/J = 5$, $V_0/J = 0.03$; $U/J = 10$, $V_0/J = 0.05$; 
$U/J = 20$, $V_0/J = 0.1$. Here and in the following, spin expectation values are plotted in units of $\hbar$.
}
\label{2d}
\end{figure}

\subsection{Results for two dimensions}
In Fig. \ref{RS2d} a typical real-space spin configuration is shown for an imbalanced two-dimensional system. Here we have chosen to label the spatial coordinates by $x$ and $z$, such that the spin direction and spatial direction can be identified.
Results of our R-DMFT calculation on the two-dimensional square lattice for the radial density and spin profiles are shown in Fig. \ref{2d}, both for the balanced system and for the situation that imbalance is induced by a nonzero chemical potential difference $\Delta \mu$. 

\subsubsection{Antiferromagnetic region}
We first turn our attention to the antiferromagnetic region in the center, where the particle density is at half-filling.
Our results show that imbalance reduces antiferromagnetic order in the center, and tilts it out of the $xy$-plane by inducing a nonzero ferromagnetic $z$-component of the spin. The effect of imbalance becomes larger with increasing interactions. This can be understood from the fact that for large $U/J$ the local spins in the insulating region interact via a Heisenberg Hamiltonian
\begin{equation}
\mathcal{H} = J_{\rm ex} \sum_{\langle i j \rangle} \hat S_i \cdot \hat S_j - \Delta \mu \sum_i \hat S_i^z .
\end{equation}
Here $J_{\rm ex} = 2 J^2/U$, 
such that with increasing $U$ the exchange coupling between the spins decreases and becomes weaker relative to the applied chemical potential difference.

\begin{figure}


\hspace{-1cm}
\includegraphics[scale = .4]{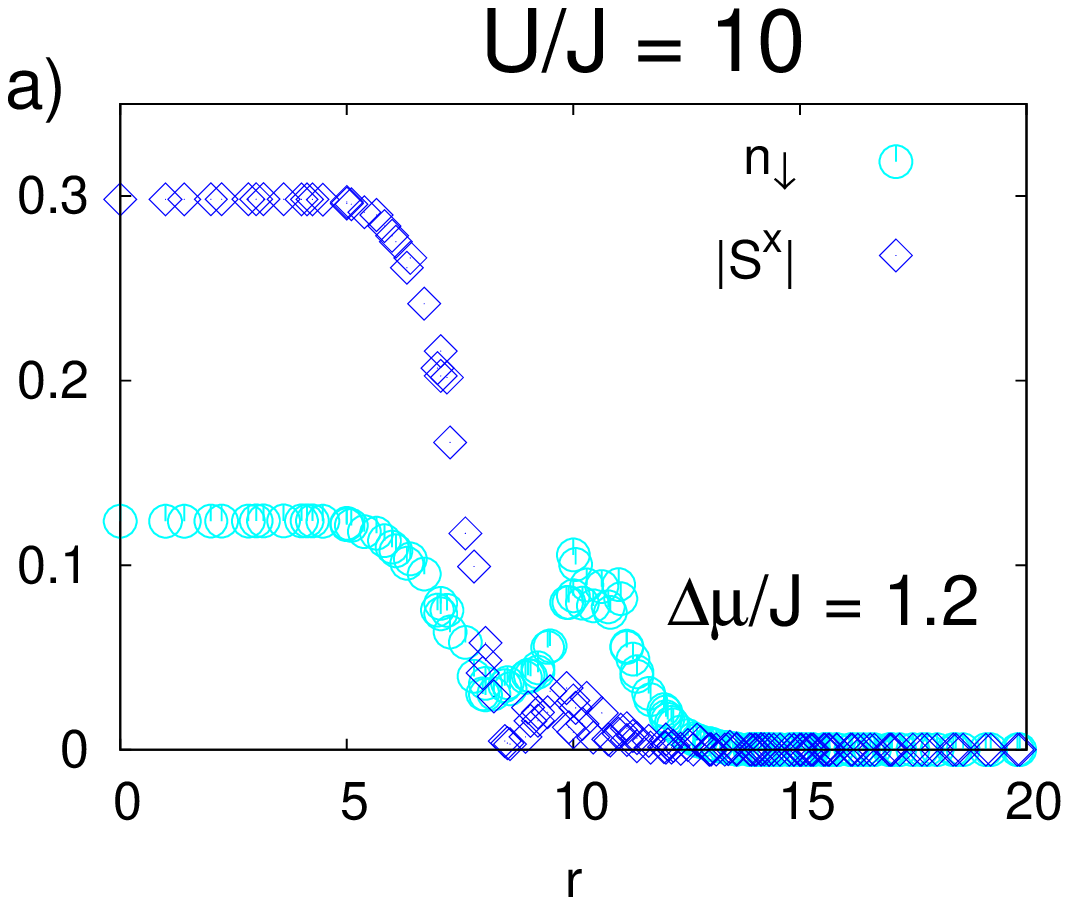}\hspace{-.8cm}
\includegraphics[scale = .4]{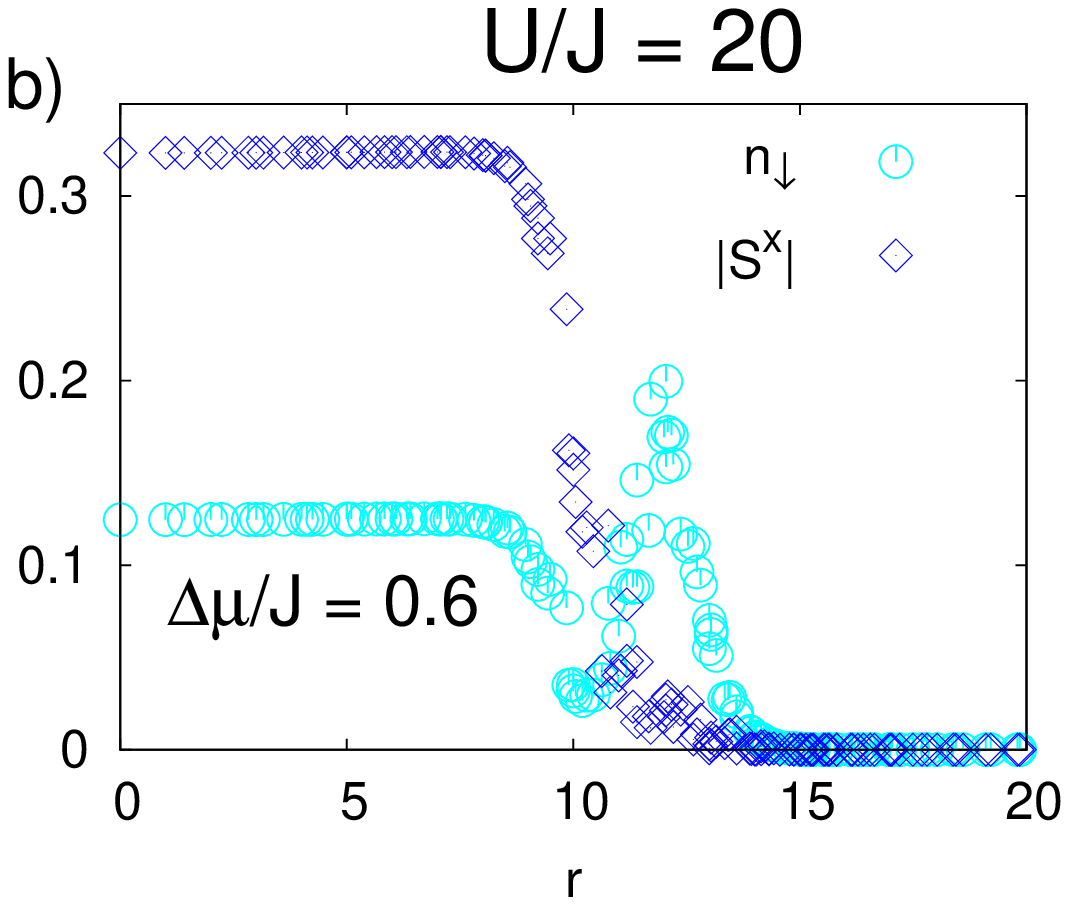}\hspace{-.5cm}

\hspace{-1cm}
\includegraphics[scale = .4]{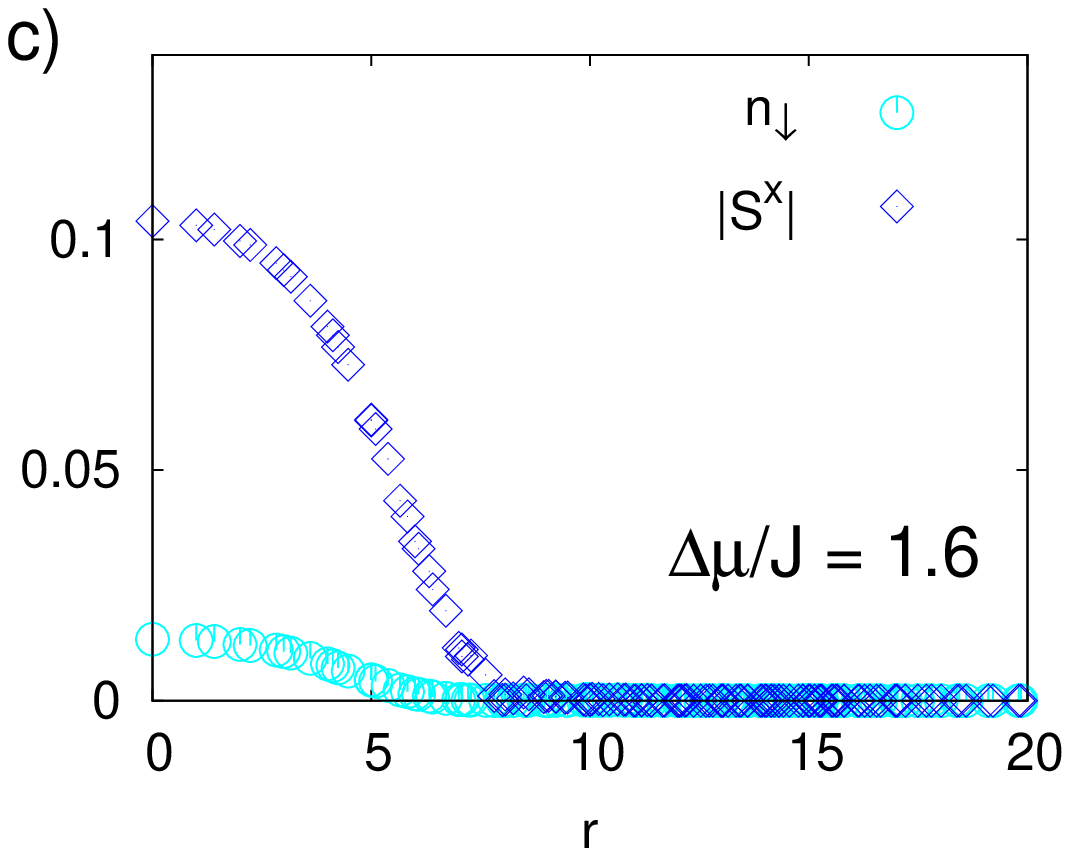}\hspace{-.8cm}
\includegraphics[scale = .4]{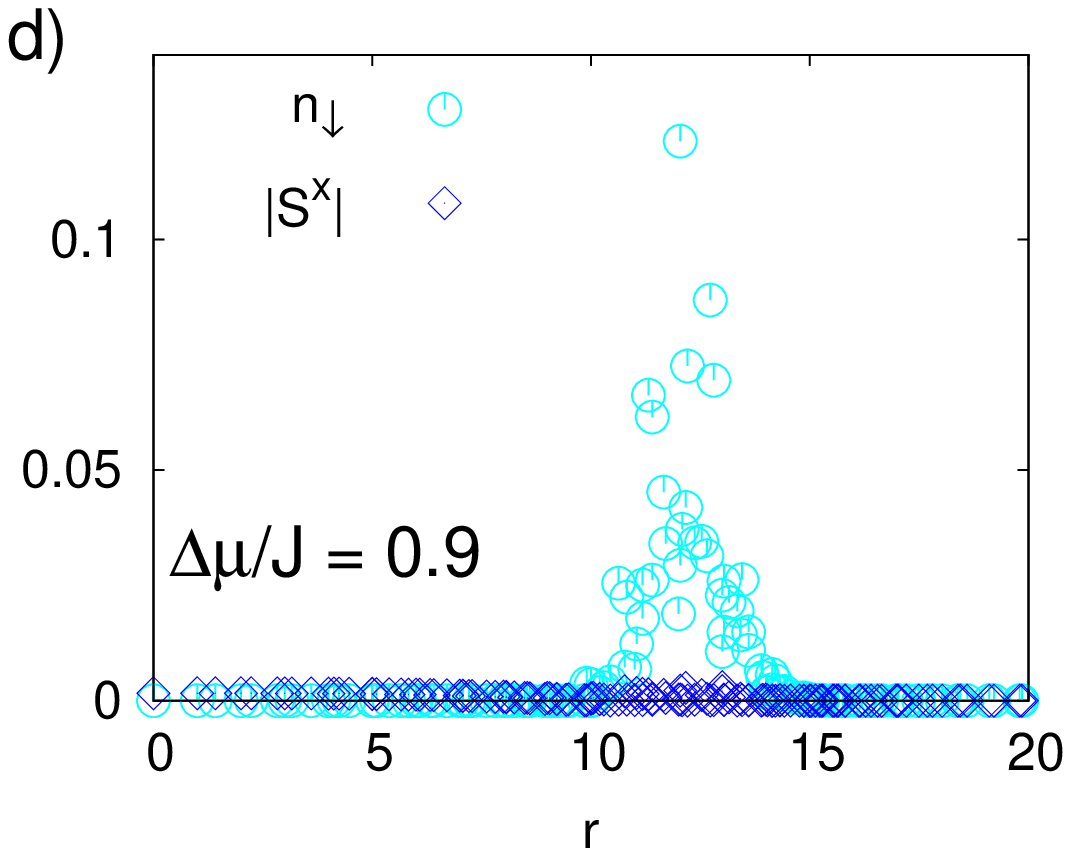}\hspace{-.5cm}

\caption{(Color online) Radial distribution for the minority density $\langle n_{i\downarrow} \rangle$ and the in-plane staggered order  $\langle \hat S^x_i \rangle$ (here plotted in absolute value) for large imbalance and different values of the repulsion: $U/J=10$ (left column) and $U/J=20$ (right column) and $\bar \mu = U/2$. Other parameters: a) $V_0/J = 0.05$, $\Delta \mu/J = 1.2$; b) $V_0/J = 0.07$, $\Delta \mu/J = 0.6$; c) $V_0/J = 0.05$, $\Delta \mu/J = 1.6$; d) $V_0/J = 0.07$, $\Delta \mu/J = 0.9$. 
}
\label{2dring}
\end{figure}

We do not observe any sign of a Stoner instability in the paramagnetic wings for the values of $U/J$ considered here, which would mean a spontaneous ferromagnetic polarization for equal chemical potentials of the spin components. In contrast, we observe that only upon the application of a finite chemical potential difference ferromagnetic order is induced in the wings.
Although we cannot establish the critical interaction above which the Stoner instability occurs, we thus find that this value is considerably shifted upwards compared with the value obtained within the Hartree Fock analysis, where spontaneous ferromagnetism was observed for even smaller values of $U/J$ than considered here \cite{Wunsch09}.  It is indeed well-known that the Hartree-Fock approximation underestimates the critical interaction for spontaneous ferromagnetism by more than an order of magnitude \cite{Vollhardt00}. Within DMFT the dynamical screening of the local repulsion is fully accounted for and only for extremely large on-site repulsion a ferromagnetic ground state was found on the homogeneous cubic lattice \cite{Obermeier97}. However, this limit is experimentally hard to reach, because the associated critical temperature  for spin order is very low. 

\subsubsection{Boundary structure}
We now turn our attention to the boundary structure between the antiferromagnetic core and the paramagnetic wings.
The results in Fig. \ref{2d} show a local maximum of the ferromagnetic polarization in the $z$-direction.  This maximum appears at the location where the antiferromagnetic order in the center disappears. The minority density shows a local minimum at this point (cf. Fig. \ref{2dring}a,b), whereas the majority density has a local maximum.  This feature was also observed in the Hartree Fock analysis \cite{Wunsch09}.  The maximum appears because the antiferromagnetic order reduces the density difference of the two components compared to the paramagnetic situation: 
 a smaller density difference leads to a larger sublattice magnetization and hence a lower energy.   
This mechanism suppresses the occurrence of phase separation in the trap center, which was found previously in the case where canted order was excluded \cite{Snoek08}. In contrast, 
the possibility of canted order allows a continuous transition between the limiting cases of an antiferromagnetic phase with equal populations of the two species and a fully polarized ferromagnetic phase in which only one of the two species is still present. 

An interesting structure emerges in the paramagnetic outer region for large imbalance: the density of minority atoms shows a second maximum (cf. Fig. \ref{2dring}) originating from the strong repulsion which pushes them to the outside. This ring-like structure is the remnant of phase separation found before \cite{Snoek08}. However, in this case it is no true phase separation, since also minority atoms are present in the central antiferromagnetic region. 
Only for large values of $U/J$ in the limit of very strong imbalance, this outer ring of minority atoms survives when the antiferromagnetic order in the center disappears, implying true phase separation. 
 In particular for $U/J=10$ we do not observe this scenario, in contrast to the case where canted antiferromagnetic order was not included \cite{Snoek08}. 
 When canted order is accounted for, the outer ring of minority atoms disappears before the antiferromagnetic order in the center vanishes, as visible in the data in Fig. \ref{2dring}a) and c).
 For $U/J=20$ still phase separation occurs, as shown Fig. \ref{2dring}b) and d): for large imbalance the antiferromagnetic order in the center breaks down, but the ring of minority atoms surrounding the phase separated central region with only majority atoms is still present. 
 Note that this phase separation cannot be identified with the Stoner instability, because it only happens for large chemical potential difference. Moverover, it means that the complete central region, including the part where the density is at half-filling, is fully polarized, whereas the minority atoms are located in a shell around this at low majority density. In contrast, the Stoner instability favors a scenario, where the region at half-filling still supports antiferromagnetic order, whereas the edges are completely polarized  \cite{Wunsch09}.

\begin{figure}
\hspace{-1cm}
\includegraphics[scale = .4]{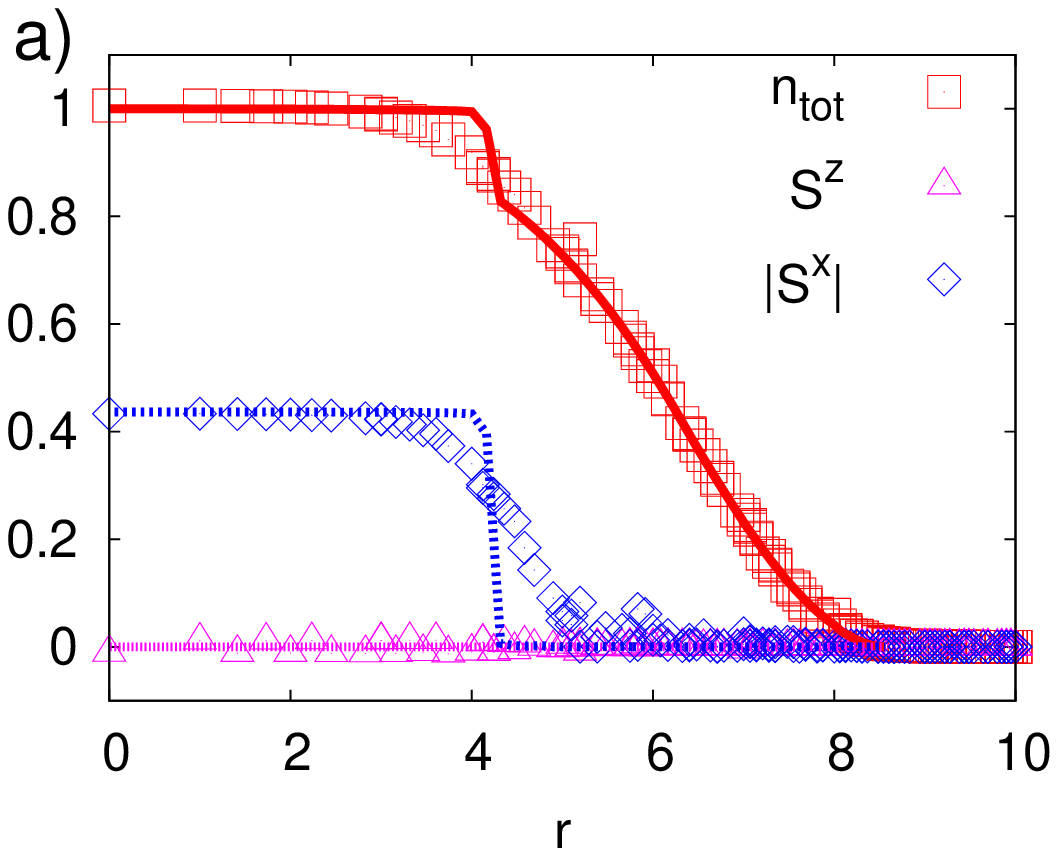}\hspace{-.8cm}
\includegraphics[scale = .4]{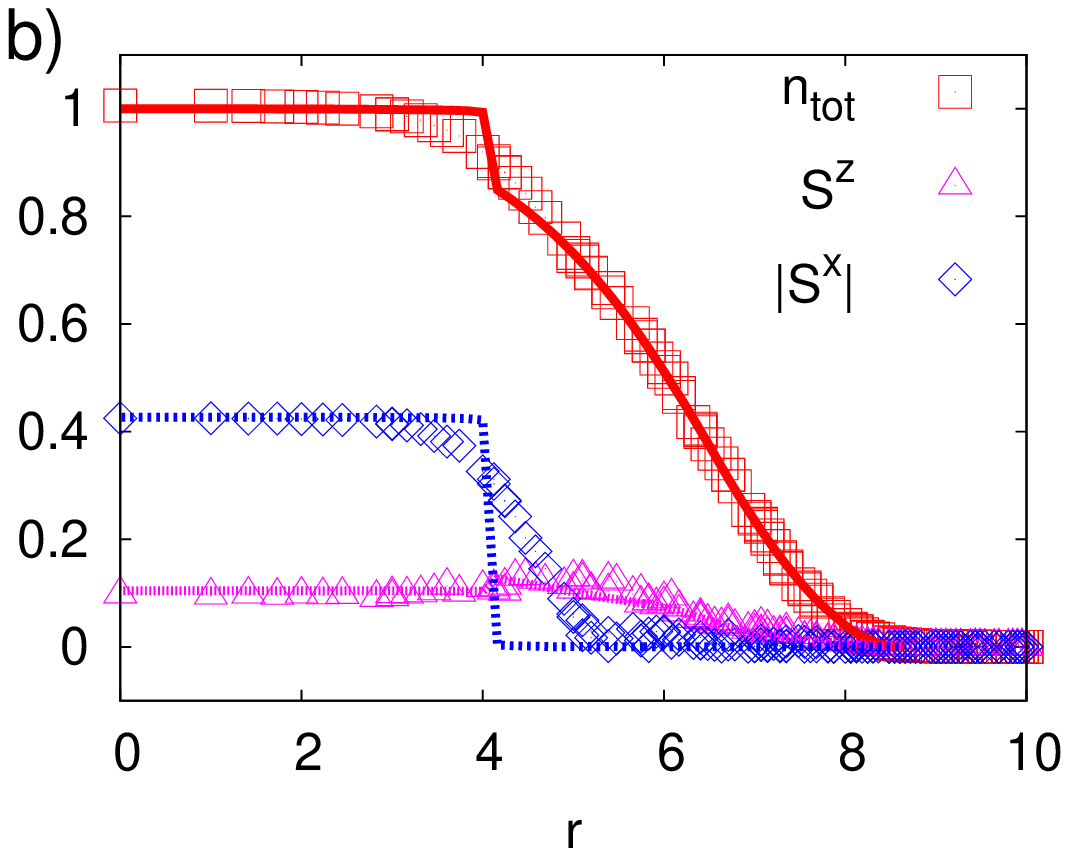}\hspace{-.5cm}

\hspace{-1cm}
\includegraphics[scale = .4]{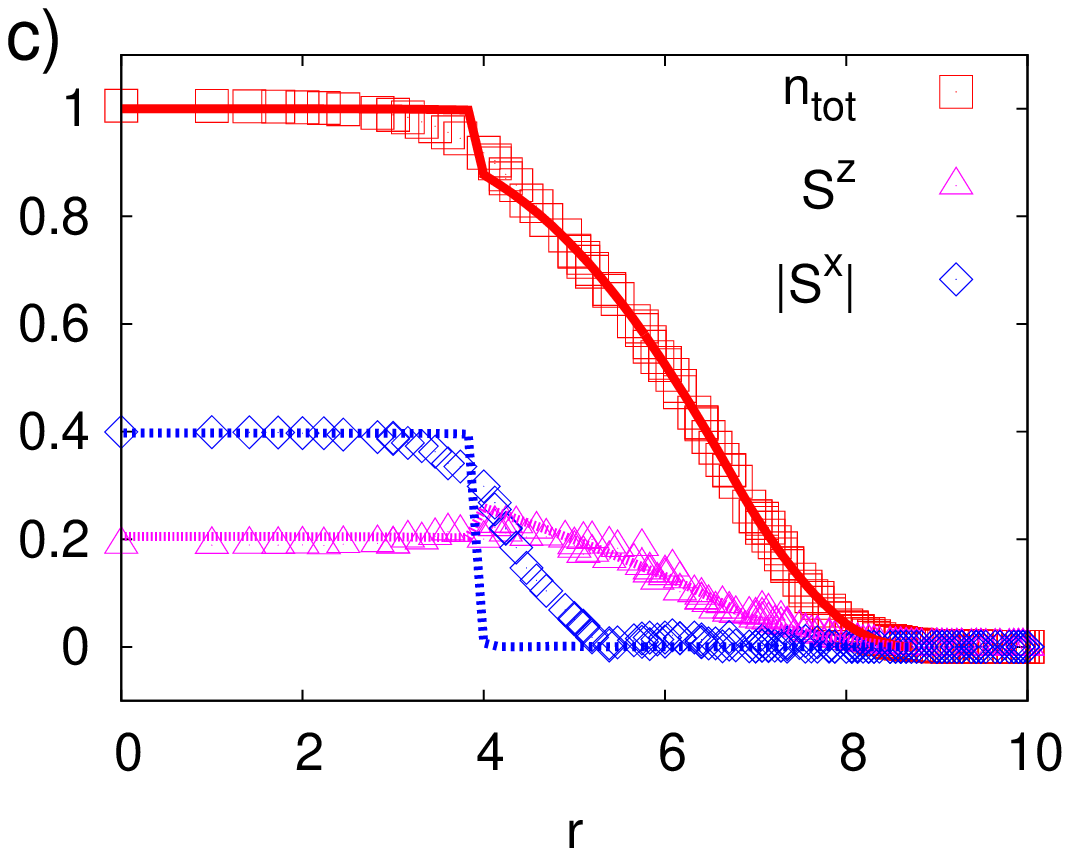}\hspace{-.8cm}
\includegraphics[scale = .4]{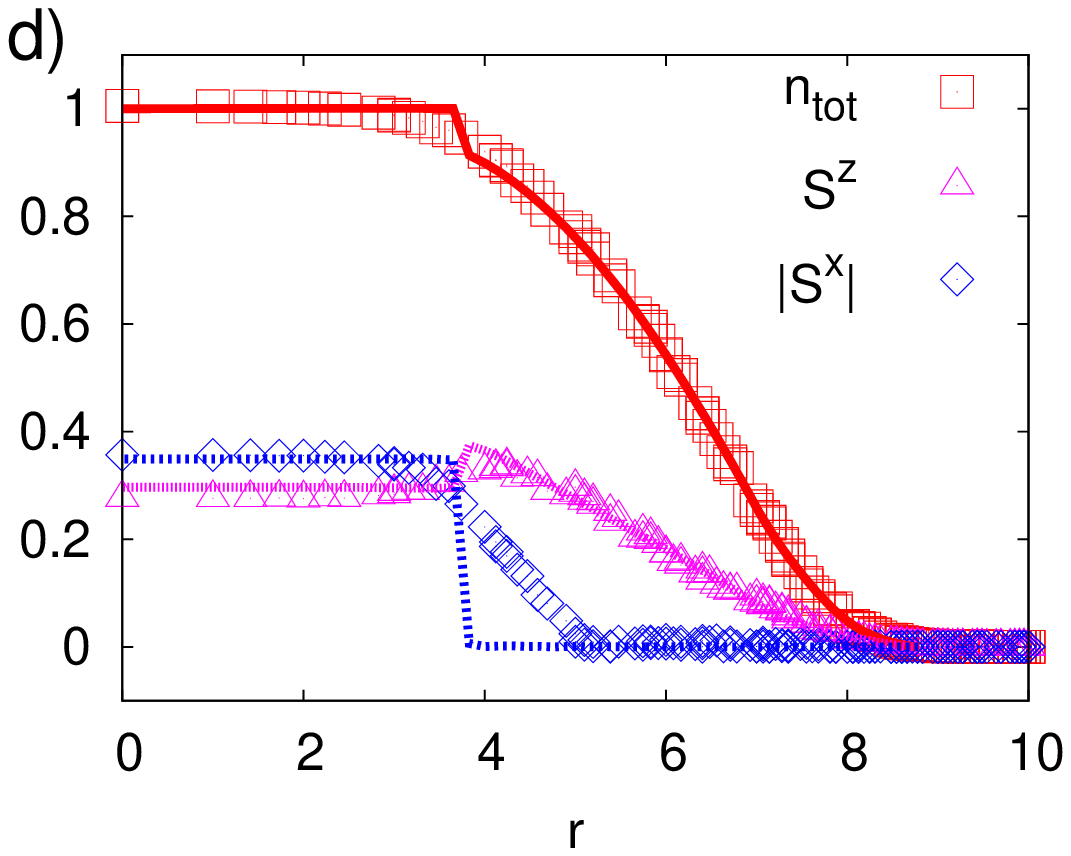}\hspace{-.5cm}

\hspace{-1cm}
\includegraphics[scale = .4]{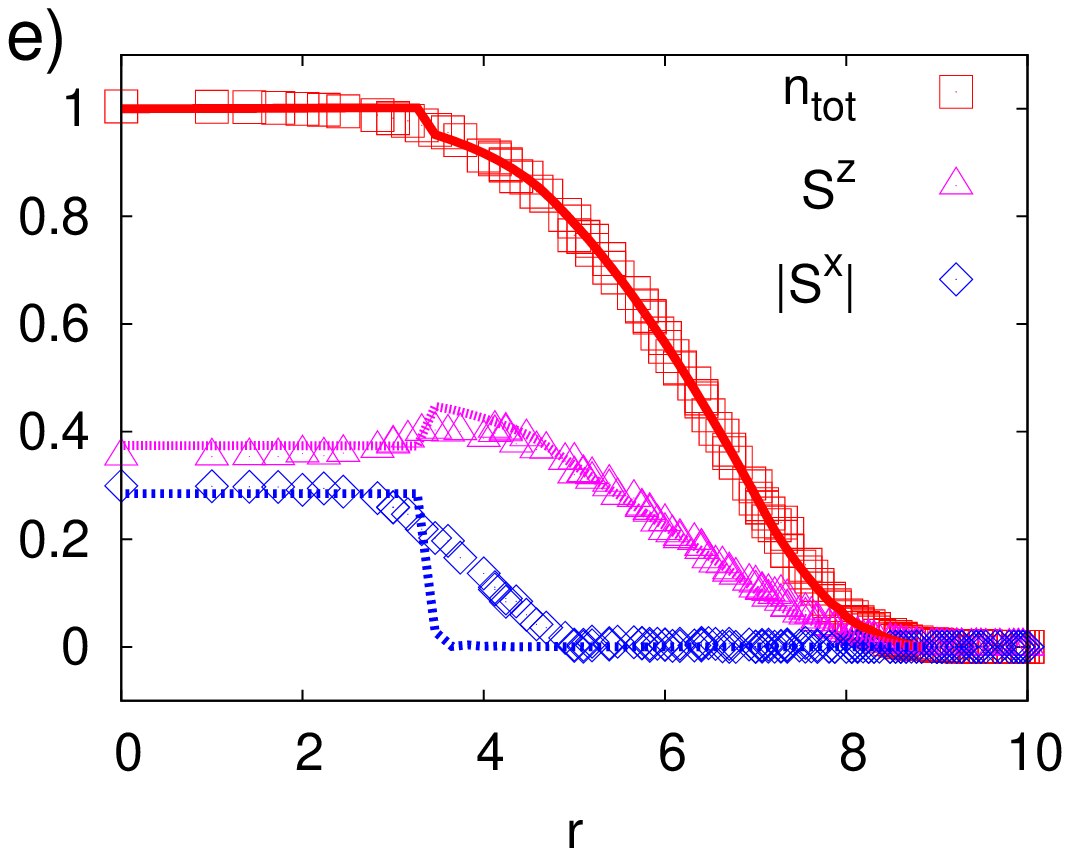}\hspace{-.8cm}
\includegraphics[scale = .4]{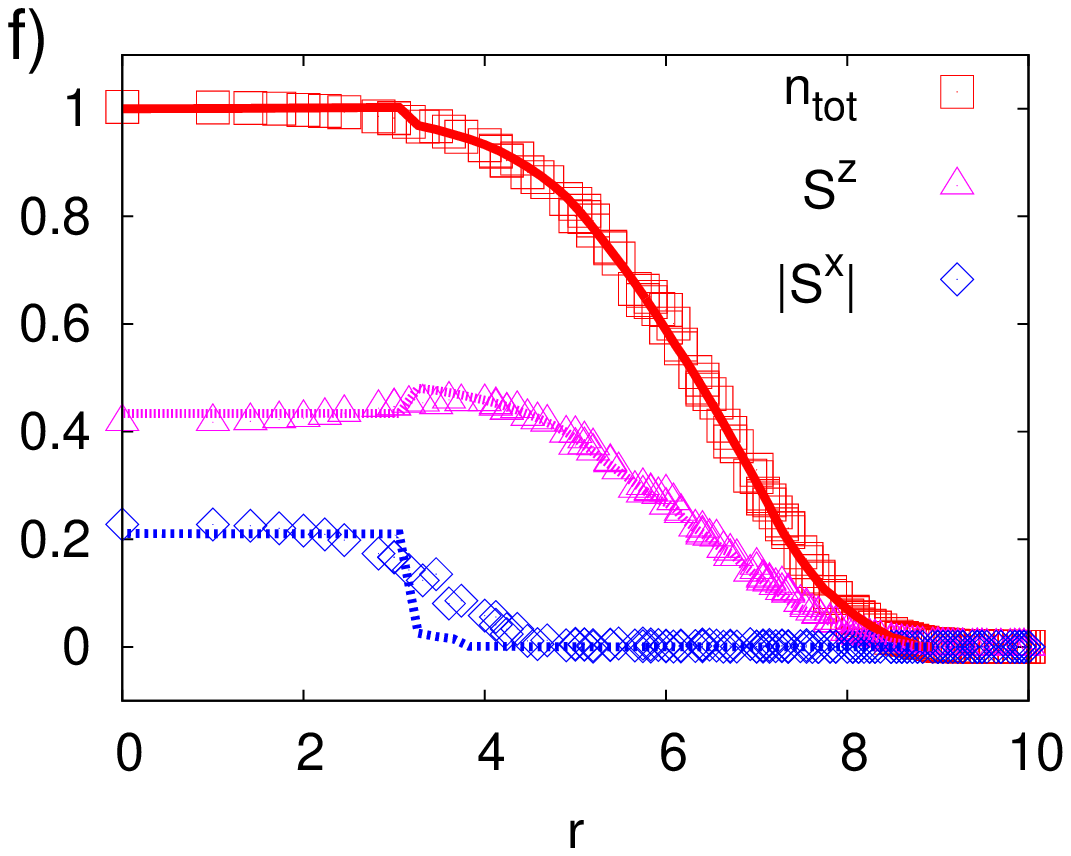}\hspace{-.5cm}

\caption{(Color online) Radial profiles for the ground state of \emph{three-dimensional} Fermi-Fermi mixtures for $U/J=10$, $\bar \mu = U/2$ and different values of $\Delta \mu$. Plotted are the total density $\langle \hat n_i \rangle = \langle \hat n_{i\uparrow} + \hat n_{i\downarrow} \rangle$, and the expectation values of the spin in $z$-direction $\langle \hat S^z_i \rangle$ and the absolute value of the expectation value in $x$-direction $|\langle \hat S^x_i \rangle|$. 
The points denote results of the full R-DMFT calculation, whereas the solid lines are obtained within the LDA (TFA) approximation combined with DMFT. 
The values for $\Delta \mu$ are: a) $\Delta \mu/J = 0$; b) $\Delta \mu/J = 0.4$; c) $\Delta \mu/J = 0.8$; d) $\Delta \mu/J = 1.2$; e) $\Delta \mu/J = 1.6$; f) $\Delta \mu/J = 2$.
The trap parameter is chosen as $V_0/J = 0.15$.}
\label{3d}
\end{figure}

\subsection{Results for three dimensions}
Results for the ground state ($T/J\to0$) in three dimensions are presented in Fig. \ref{3d}. We compare the full R-DMFT results with a local density approximation (LDA) (previously also denoted as a Thomas-Fermi approximation (TFA)\cite{Snoek08}),  where the trap is modeled within DMFT by a locally varying chemical potential combined with the homogeneous density of states of the cubic lattice. In order to facilitate the comparison, the absolute value of the staggered spin expectation value in $x$-direction is plotted in Fig. \ref{3d}. This makes it impossible to distinguish ferromagnetic and anti\-ferromagnetic order in the figure; however, as in Fig. \ref{2d}  the in-plane $xy$-spin order described by $S^x$ is always staggered.
For the total density we observe good agreement between the LDA+DMFT and the full R-DMFT results; the only difference being that the LDA results show a small discontinuity, which is smoothened in the R-DMFT calculation. 
In contrast, the agreement between the LDA and R-DMFT results for the spin order parameters is far less good, as also observed for the case of balanced mixtures \cite{Snoek08, Gorelik10}. In particular, antiferromagnetic spin order extends much further into the region with total density lower than half-filling (i.e. $n_i<1$) than LDA analysis predicts. This is a consequence of a proximity effect of the antiferromagnetic insulator close to the paramagnetic boundary layer.

Examining the numerical results in more detail we observe similar features as for the two-dimensional case in Fig. \ref{2d}. 
 A maximum of the ferromagnetic polarization in the $z$-direction is also visible in the data in Fig. \ref{3d}, although it is more pronounced in the LDA curve than in the R-DMFT data, and in general less pronounced than in two dimensions. This is because the ratio of the interaction to the band-width $U/2zJ$ ($z$ being the number of neighbors) is chosen smaller here. This is also the reason that phase separation is not observed for the case of strong imbalance, even though large values of $\Delta \mu$ are considered. The minority atoms still form a shell for large imbalance, but this shell disappears before the antiferromagnetic order in the center is destroyed.
 In order to observe  phase separation, a stronger repulsion would be  needed. 
Also, a Stoner instability is not observed in the three dimensional case for the value of the repulsion chosen in this case.  

\begin{figure}
\begin{center}
\includegraphics[width = \columnwidth]{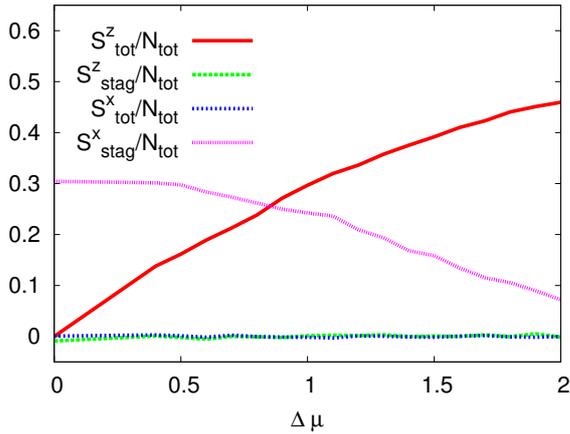}
\end{center}
\caption{(Color online) Global observables for the three-dimensional system obtained by R-DMFT at $U/J = 10$, $\bar \mu = U/2$ and $V_0/J = 0.15$ as a function of the chemical potential difference $\Delta \mu$. Shown are the total magnetization in $x$ and $z$-direction $S^{x,z}_{\rm tot}$ and the staggered magnetization in $x$ and $z$-direction $S^{x,z}_{\rm stag}$, all normalized to the particle number $N_{\rm tot}$ }
\label{3ddata}
\end{figure}

Our data in three dimensions are summarized in Fig. \ref{3ddata}. Here we plot global observables as a function of the chemical potential difference: the total magnetization in $x$ and $z$ direction $S^{x,z}_{\rm tot} = \sum_i \langle \hat S^{x,z}_i \rangle$ and the staggered magnetization in $x$ and $z$ direction $S^{x,z}_{\rm stag} = \sum_i (-1)^{i_x + i_y} \langle \hat S^{x,z}_i \rangle$, normalized to the total particle number $N_{\rm tot} = \sum_i \langle \hat n_{i\uparrow} + \hat n_{i\downarrow} \rangle$. Fig \ref{3ddata} shows that the total magnetization always points in the $z$-direction; which is also the direction in which it is induced experimentally via the spin imbalance. This means that the system is not frustrated and, for the parameters considered here, the interesting scenarios like the one pursued in Ref. \cite{Wunsch09} are not realized.

\section{Conclusions}
We have solved the fermionic Hubbard model with repulsive interactions in a harmonic trap by means of Real-Space Dynamical Mean-Field Theory, with a focus on the ground state properties of imbalanced spin populations. We considered systems with a density of one particle per lattice site in the center, for which a shell structure appears: an insulating regime in the center, in which antiferromagnetic order can arise, is surrounded by a Fermi liquid regime without magnetic order. Imbalance leads to canted antiferromagnetic order in the center, with an interesting boundary structure to the paramagnetic edge, which we investigated in detail. 

Due to the possibility of canting, the antiferromagnetism turns out to be very stable against imbalance:  only for sufficiently large values of the repulsion and at large imbalance 
phase separation occurs. In this regime the central region becomes completely polarized: only the majority component is present and forms a band insulator in the center.  The minority atoms organize themselves in a shell around this central plateau.
  
We do not observe a Stoner instability toward spontaneous ferromagnetic order for the moderately large values of the repulsion considered here. The critical interaction to observe this phenomenon is thus necessarily relatively large, leading to a small critical temperature for ferromagnetic order. 

Canted antiferromagnetic order is more challenging to detect experimentally than antiferromagnetic order in the $z$-direction, because many proposed detection schemes are only sensitive to staggered order which is diagonal in the basis of the physical particles constituting the system. 
 This limitation does, however, not apply to Bragg scattering\cite{corcovilos10}, which is therefore the experimental method of choice to detect canted antiferromagnetic order.

\acknowledgements

We thank I. Bloch, E. Demler, W. Ketterle, A. Koetsier and H.T.C. Stoof for useful discussions. This 
work was supported by the Nederlandse Organisatie voor Wetenschappelijk Onderzoek (NWO) and the German Science Foundation 
DFG via Forschergruppe FOR 801, Sonderforschungsbereich SFB-TRR 49 and the DIP project BL 574/10-1.



\newcommand{\PRL}{Phys.\ Rev.\ Lett.}
\newcommand{\PRB}{Phys.\ Rev.\ B}
 \newcommand{\PRA}{Phys.\ Rev.\ A}
 \newcommand{\RMP}{Rev.\ Mod.\ Phys.}

\end{document}